\documentclass[aps,twocolumn,prl,longbilbliography,superscriptaddress,shopacs]{revtex4-2} 

\usepackage{graphicx, epstopdf}
\usepackage{bm}
\usepackage[none]{hyphenat}
\usepackage[colorlinks=true]{hyperref} 
\usepackage{multirow, tabularx}
\usepackage{float,xcolor}
\usepackage{xr}
\usepackage{amsmath, cancel}
\usepackage{placeins}
\usepackage{float}
\usepackage{soul}
\usepackage{ulem}

\newcommand{\FeNbS}[1]{\text{Fe$_\text{#1}$NbS$_2$}}
\newcommand{\eg}{$e_{g}$}
\newcommand{\ttg}{$t_{2g}$}

\begin{document}



\title{Electronic origin of delicate antiferromagnetism in \FeNbS{x}}

\author{Wenxin Li}
    \affiliation{Department of Applied Physics, Yale University, New Haven, Connecticut 06511, USA}
\author{Jonathan T. Reichanadter}
    \affiliation{Department of Electrical Engineering, University of California Berkeley, California, 94720, USA}
    \affiliation{Department of Physics, University of California Berkeley, California, 94720, USA}
    \affiliation{Material Sciences Division, Lawrence Berkeley National Lab, Berkeley, California, 94720, USA}
\author{Shan Wu}
    \affiliation{Department of Physics, University of California Berkeley, California, 94720, USA}
    \affiliation{Material Sciences Division, Lawrence Berkeley National Lab, Berkeley, California, 94720, USA}
    \affiliation{Department of Physics, Santa Clara University, Santa Clara, CA, 95053}
\author{Ji Seop Oh}
    \affiliation{Department of Physics, University of California Berkeley, California, 94720, USA}
    \affiliation{Department of Physics and Astronomy, Rice University, Houston, Texas 77024, USA}
    \affiliation{Department of Applied Physics, Sookmyung Women’s University, Seoul 04310, Republic of Korea}
    \affiliation{Institute of Advanced Materials and Systems, Sookmyung Women’s University, Seoul 04310, Republic of Korea}
\author{Rourav Basak}
    \affiliation{Department of Physics, University of California San Diego, California, 92093, USA}
\author{Shannon C. Haley}
    \affiliation{Department of Physics, University of California Berkeley, California, 94720, USA}
\author{Siqi Wang}
    \affiliation{Department of Applied Physics, Yale University, New Haven, Connecticut 06511, USA}
\author{Joshua E. Chaparro Mata}
    \affiliation{Department of Applied Physics, Yale University, New Haven, Connecticut 06511, USA}
\author{Elio Vescovo}
    \affiliation{National Synchrotron Light Source II, Brookhaven National Laboratory, Upton, New York 11973, USA}
\author{Donghui Lu}
    \affiliation{Stanford Synchrotron Radiation Lightsource, SLAC National Accelerator Laboratory, Menlo Park, California 94025, USA}
\author{Makoto Hashimoto}
    \affiliation{Stanford Synchrotron Radiation Lightsource, SLAC National Accelerator Laboratory, Menlo Park, California 94025, USA}
\author{Christoph Klewe}
    \affiliation{Advanced Light Source, Lawrence Berkeley National Laboratory, Berkeley, California 94720, USA}
\author{Suchismita Sarker}
    \affiliation{Cornell High Energy Synchrotron Source, Cornell University, Ithaca, New York 14853, USA}
\author{Jessica L. McChesney}
    \affiliation{Advanced Photon Source, Argonne National Laboratory, Lemont, Illinois 60439, USA}
\author{Alex Fra\~n\'o}
    \affiliation{Department of Physics, University of California San Diego, California, 92093, USA}
\author{James G. Analytis}
    \affiliation{Department of Physics, University of California Berkeley, California, 94720, USA}
    \affiliation{CIFAR Quantum Materials, CIFAR, Toronto, Ontario M5G 1M1, Canada}
\author{Robert J. Birgeneau}
    \affiliation{Department of Physics, University of California Berkeley, California, 94720, USA}
    \affiliation{Material Sciences Division, Lawrence Berkeley National Lab, Berkeley, California, 94720, USA}
\author{Jeffrey B. Neaton}
    \affiliation{Department of Physics, University of California Berkeley, California, 94720, USA}
    \affiliation{Material Sciences Division, Lawrence Berkeley National Lab, Berkeley, California, 94720, USA}
    \affiliation{Kavli Energy Nanosciences Institute at Berkeley, Berkeley, California, 94720, USA}
\author{Yu He}
    \email{yu.he@yale.edu}
    \affiliation{Department of Applied Physics, Yale University, New Haven, Connecticut 06511, USA}

\date{\today}


\begin{abstract}
Among the family of intercalated transition-metal dichalcogenides (TMDs), Fe$_{x}$NbS$_2$ is found to possess unique current-induced resistive switching behaviors, tunable antiferromagnetic states, and a commensurate charge order, all of which are tied to a critical Fe doping of $x_c$~=~1/3.
However, the electronic origin of such extreme stoichiometry sensitivities remains unclear. 
Combining angle-resolved photoemission spectroscopy (ARPES) with density functional theory (DFT) calculations, we identify and characterize a dramatic eV-scale electronic restructuring that occurs across the $x_c$. Moment-carrying Fe 3$d_{z^2}$ electrons manifest as narrow bands within 200 meV of the Fermi level, distinct from other transition metal intercalated TMD magnets. These states strongly hybridize with itinerant electrons in TMD layer, rapidly lose coherence above $x_c$ due to correlation-driven effects. This sudden quasiparticle decoherence collapses the Fe-Nb hybridization, which explicitly suppresses the out-of-plane effective Fe-Fe exchange interaction, driving the transformation of the magnetic ground state from an antiferromagnetic stripe phase to a zigzag phase.
These observations resemble the exceptional electronic and magnetic sensitivity of strongly correlated systems, and demonstrate that quantifying orbital-specific hybridization via ARPES offers an alternative pathway to evaluate effective magnetic exchange in metallic magnets, complementing inelastic neutron and resonant x-ray scattering probes.
\end{abstract}

\maketitle

\begin{figure}[!tbh]
\includegraphics[width=0.9\columnwidth,clip,angle =0]{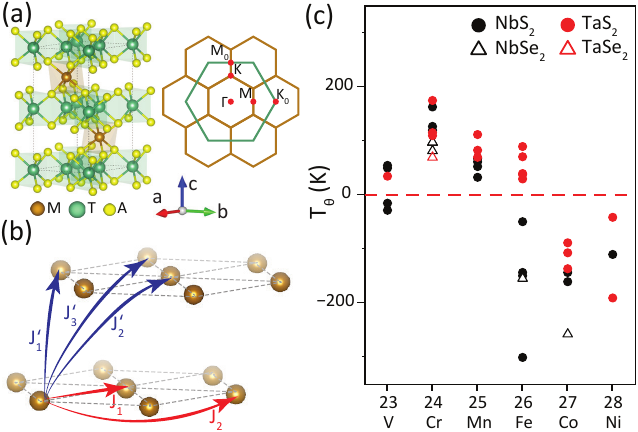}
\caption{\label{fig1}
    Structural and magnetic property of intercalated TMD magnets. (a) The crystallographic structure and Brillouin zone of 3$d$ metal intercalated TMDs M$_x$TA$_2$ with $x$~=~1/3.
    Brown and Green hexagonal traces circumscribe the in-plane Brillouin zones for M$_{1/3}$TA$_2$ and TA$_2$ respectively. (b) Depiction of the Heisenberg exchange terms between in-plane and out-of-plane nearest, next-nearest, and third-nearest neighbor intercalant ions in M$_{x}$TA$_2$.
    (c) Compilation of the Curie-Weiss temperatures for various M$_x$TA$_2$ systems with $x$~=~1/3~\cite{TMD1968,TMD1970,TMD1971,TMD1975,TMD1976, parkin1980,ghimire2018,VNbS_1,VNbS_2,CrTaS,CrNbSe_1,CrNbSe_2,MnTaS,FeNbS_CW,FeTaS_1,FeTaS_2,CoNbS,CoTaS,NiNbTaS}.
}
\end{figure}

\noindent
Layered transition-metal dichalcogenides (TMD) TA$_2$ (T = Ta, Nb, Mo; A = Se, S) constitute a family of materials that are pivotal to modern condensed matter physics research, owing to their rich physical properties, ranging from charge density wave to superconductivity~\cite{wilson1974,borisenko2009,valla2004,rahn2012}.
While most TMDs assume a weakly correlated non-magnetic ground state in their pristine phase~\cite{reviewTMD1,reviewTMD2}, their layered crystal structure often allows 3$d$ transition metal intercalants to populate the van der Waals (vdW) gap (M$_x$TA$_2$, M = 3$d$ transition metal). This leads to a versatile class of tunable layered magnets that are often also highly metallic~\cite{parkin1980,friend1977}, where indirect exchange via itinerant electrons is considered indispensable given the large M-site distances.
Among all possible intercalant concentrations, materials with intercalation ratio $x_c$~=~1/3, where the intercalated atoms typically form into a $\sqrt3 \times \sqrt3$ superlattice~\cite{boswell1978} with non-centrosymmetric space group $P6_3 22$ and a folded Brillouin zone (Fig.~\ref{fig1}~(a)), have drawn considerable interest with their extensive display of diverse magnetic and electronic properties \cite{togawa2012,braam2015,Kousaka_2016,kousaka2009,karna2019, togawa2015,ghimire2018,aoki2019,Park2022}.

\begin{figure*}[!tbh]
\includegraphics[width=1.8\columnwidth,clip,angle =0]{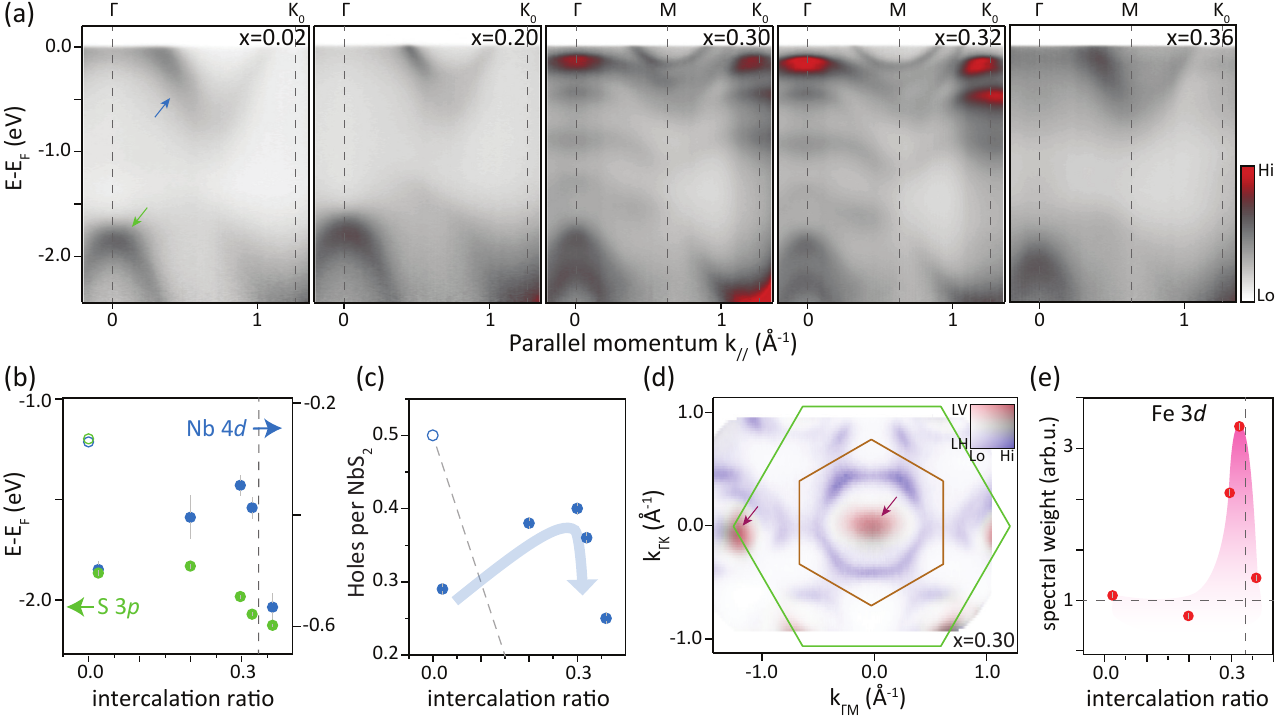}
\caption{\label{fig2}
ARPES measurement of Fe$_{x}$NbS$_2$ across different Fe stoichiometry.
    (a) Energy-momentum cut along $\Gamma$-$M$-$K_0$ with different Fe stoichiometry.
    (b, c) Evolution of the position of S 3$p$ and Nb 4$d$ derived bands and number of holes calculated using Luttinger theorem with increasing Fe stoichiometry. Data for $x$ = 0 (hollow points) corresponds to $k_z$ = 0 and are extracted from Fig.~1~(c) (DFT calculated band structure) of~\cite{NbS2ARPES}.
    (d) Constant energy contours integrated between $E_B$ = 0.2 eV and 0.1 eV for $x$ = 0.30. Purple arrows indicate Fe-derived bands. Axis labels $k_{\Gamma M}$ and $k_{\Gamma K}$ mark directions to high-symmetry points in the Fe$_{1/3}$NbS$_2$ Brillouin zone.
    (e) Evolution of spectral weight of the Fe-derived bands with increasing Fe stoichiometry. Spectral weight integrated between [0,1] eV binding energy and $\pm 0.2$~\AA$^{-1}$~of $\Gamma$.
}
\end{figure*}

\noindent
Figure~\ref{fig1} (c) tallies the Curie-Weiss temperature $T_\theta$ for different M$_x$TA$_2$ materials with highly tunable magnetic correlations~\cite{TMD1968,TMD1970,TMD1971,TMD1975,TMD1976, parkin1980,ghimire2018,VNbS_1,VNbS_2,CrTaS,CrNbSe_1,CrNbSe_2,MnTaS,FeNbS_CW,FeTaS_1,FeTaS_2,CoNbS,CoTaS,NiNbTaS}.
Except for V$_{1/3}$NbS$_2$, which has a controversial magnetic ground state and was recently proposed as an altermagnet~\cite{altermag_1,altermag_2}, Fe intercalation is uniquely associated with both ferromagnetic ($T_\theta>0$) and antiferromagnetic ($T_\theta<0$) correlations depending on the atomic species of the TMD layers.
This implies an active role of the TMD layer in mediating long-range Fe-Fe magnetic exchange, and makes Fe-intercalated TA$_2$ materials hosts for highly tunable magnetism and magnetoelectric transport properties. 
Recent neutron scattering and muon-spin relaxation ($\mu$SR) measurements reported two distinct magnetic phases across $x_c$~=~1/3, i.e., the antiferromagnetic stripe order at $k_{m1}$ = (0.5, 0, 0) for $x <$ $x_c$, and the antiferromagnetic zigzag order at $k_{m2}$ = (0.25, 0.5, 0) for $x >$ $x_c$~\cite{sw_prx2022,FNS_usr}.
These nearly degenerate magnetic phases in Fe$_{x}$NbS$_2$ are sensitively decided by next-nearest-neighbor in-plane ($J_2$) versus out-of-plane ($J_2^\prime$) exchange interactions [Fig.~\ref{fig1}~(b)]~\cite{sw_prx2022} as suggested from a Heisenberg Hamiltonian model~\cite{weber2021}, and are believed to cause its current-induced resistance switching property~\cite{nair2020,maniv2021_1,maniv2021_2}.
However, determining these Heisenberg exchange terms via the traditional approach of linear spin wave fits of inelastic neutron or x-ray scattering spectra often faces challenges in metallic magnets like Fe$_{x}$NbS$_2$ due to excessive free-carrier damping~\cite{Landaudamping-1,Landaudamping-2,Landaudamping-3}, highlighting the need for a complementary route.
Moreover, a commensurate three-dimensional charge order, concomitant with the magnetic ordering, was recently discovered for $x>x_c$, signaling strongly coupled charge and magnetic degrees of freedom~\cite{CO-PRL}.
\\

\noindent
It remains a mystery why both the electrical transport and magnetic ordering properties experience a sudden change across $x_c$ = 1/3 in and only in Fe$_x$NbS$_2$.
Previous electronic structure investigations in various intercalated TMD systems show that the low-energy electronic structure mainly derives from the TMD layers, while the charge doping and minor band hybridization effects come from the intercalated 3$d$ transition metal~\cite{TMD_rigidband,Battaglia2007_MnnbS,Garb2022_CoNbS,Qin2022_CrNbS,Sirica2016_CrNbS,Xie2023,PDCking2023,NiNbS2_band,Santi2025_FeTaS2band,EdwaresFNS,FeCoNbS,newFeNbSe}.
While previous electronic structure studies on Fe$_{1/3}$NbS$_2$~\cite{EdwaresFNS}~\cite{FeCoNbS} and other TM-intercalated dichalcogenides (including but not limited to Co$_{1/3}$NbS$_2$~\cite{Garb2022_CoNbS}, Cr$_{1/3}$NbS$_2$~\cite{Sirica2016_CrNbS}, Ni$_{1/3}$NbS$_2$~\cite{NiNbS2_band}, (Fe, Co)$_{1/3}$TaS$_2$~\cite{Santi2025_FeTaS2band} and Fe$_{1/3}$NbSe$_2$~\cite{newFeNbSe})
identify emergent electronic states, the precise orbital link between these features and the intercalant species remains obscure. Crucially, the sensitivity of doping-controlled electronic and magnetic properties, i.e., how the transition between distinct magnetic ground states emerges from the underlying electronic structure, has not been explored,
motivating systematic electronic and structural investigations into the roles of Fe intercalants, especially whether and how low-energy itinerant electrons can affect the delicate indirect Fe-Fe exchange.
Understanding this requires the use of electronic structure probes to identify electronic states that carry moments and states that mediate magnetic exchange.
In this work, we combine angle-resolved photoemission spectroscopy (ARPES), X-ray photoelectron spectroscopy (XPS), single crystal X-ray diffraction (XRD), X-ray absorption spectroscopy (XAS) and density functional theory (DFT) calculations to investigate the electronic structure of Fe$_x$NbS$_2$ near the critical doping $x_c$, with a specific focus on how the interaction between itinerant and local electrons help provide insights into effective exchange interactions.
\\


\noindent
Given the extreme surface sensitivity of photoemission experiments~\cite{ARPESreview2003,ARPESreview2021} and the existence of inhomogeneous surface termination in many intercalated TMD systems~\cite{PDCking2023,Xie2023}, in particular, the putative surface termination effect in Fe$_{1/3}$NbS$_2$ reported in~\cite{EdwaresFNS}, micro-spot XPS measurements are performed on \textit{in-situ} cleaved bulk sample surfaces to examine the iron homogeneity. Compared to the micron-scale inhomogeneity reported in V$_{1/3}$NbS$_2$~\cite{PDCking2023} and Cr$_{1/3}$NbS$_2$~\cite{Xie2023}, the surface iron inhomogeneity here can be constrained to less than 2.8\% over similar length scales (see Fig.~\ref{SI-SI_XPS} in Supplementary Information (SI)~\cite{SI}), enabling systematic electronic structure investigation as the bulk doping is tuned across $x_c$ = 1/3.
Figure~\ref{fig2} (a) shows the electronic structure measured with ARPES along $\Gamma$-$K_0$ from $x$~=~0.02 to $x$~=~0.36 (see Fig.~\ref{SI-SI_EDX} for spatial composition analysis~\cite{SI}).
For doping close to $x_c$, the $\sqrt3 \times \sqrt3$ Fe-superlattice forms (Fig.~\ref{SI-SI_XRD}~\cite{SI}) so that the $M$ point of the Brillouin zone of Fe$_{1/3}$NbS$_2$ lies in the middle of $\Gamma$ and $K_0$ points.
\\

\noindent
We first investigate the charge doping effect introduced by Fe intercalation. By tracking the energies of the band extrema for the Nb 4$d$ and the S 3$p$ bands, along with the Luttinger volume of the Nb 4$d$ hole pockets around $\Gamma$ and $K$~\cite{Luttinger} (see Fig.~\ref{SI-SI_EDC}~(a-c) for detailed procedures~\cite{SI}), Fig.~\ref{fig2}~(b) and (c) shows that Fe intercalants mediate substantial charge transfer at very dilute doping ($x=0.02$) consistent with a rigid-band model \cite{parkin1980,TMD_rigidband}.
The Nb 4$d$ and S 3$p$ bands experience the same energy shift, indicating uniform electron doping to the TMD layer. Assuming full Fe ionization to Fe$^{2+}$~\cite{sw_prx2022}, the charge transfer at $x=0.02$ is larger than expected (gray dashed line in Fig.~\ref{fig2}~(c)), which may be attributed to intrinsic tendency for excess Nb formation (Nb$_{1+\delta}$S$_2$) in the NbS$_2$ part of the sample~\cite{NbS2DFT} (see Fig.~\ref{SI-SI_EDX} and related discussion in SI~\cite{SI}).
Further iron intercalation causes this charge transfer process to reverse course, where the Nb 4$d$ bands are gradually depleted (rising band bottom accompanied by a decreasing Luttinger volume), indicating potential electron localization at Fe sites.
Only when doped to $x > x_c$, electrons move to fill Nb 4$d$ bands again.
On the other hand, S 3$p$ bands shift toward higher binding energy with increasing Fe intercalation, affirming the presence of low-energy Fe state while increasing the global chemical potential.
This shift of the S 3$p$ bands further suggests reversed charge flow from Nb 4$d$ states to Fe 3$d$ states due to strong Fe-S bonds as the system approaches $x_c$.

\noindent
This non-monotonic evolution in charge transfer behavior naturally raises the question of where the extra Fe 3$d$ electrons go, which by Hund's rules carry a net magnetic moment in a high spin state according to prior neutron scattering studies~\cite{sw_prx2022}.
Strikingly, at $x$ = 0.30 and $x$ = 0.32 where Nb states receive the least charge doping, a cascade of low-energy, weakly dispersive bands form at $\Gamma$, with in-plane bandwidths less than 150 meV. 
The Fe 3$d$ character of these states are confirmed via Resonant Photoemission Spectroscopy (RPES) measurements at Fe $L$ edge (Fig.~\ref{SI-SI_XAS}~\cite{SI}).
Figure~\ref{fig2}~(d) shows the momentum-dependent spectral intensity integrated between $E_B$ = 0.1 eV and 0.2 eV where the Fe 3$d$ bands are primarily located. Here, orbital contributions reflected through orthogonal incident photon polarizations are color-coded in red and blue~\cite{2Dcolor}.
Via photoemission matrix element analysis (Fig.~\ref{SI-SI_matrix}~\cite{SI}), these low-energy narrow bands are mainly of 3$d_{z^2}$ orbital character~\cite{matrix}, and are found to repeat corresponding to the Fe-sublattice Brillouin zone. They belong to the \ttg \ states in the Fe ion's local symmetry coordinate basis, consistent with our spin-resolved density functional theory (DFT) calculations (Fig.~\ref{SI-SI_pdos_zigzag}~\cite{SI}), and are expected to carry local moments despite their very low binding energy~—~a stark contrast to what is typically observed in other 3$d$ metal-intercalated TMDs~\cite{Garb2022_CoNbS,Sirica2016_CrNbS,NiNbS2_band,Santi2025_FeTaS2band,newFeNbSe,CoNbSDFT,CrNbSDFT}.
This not only offers a natural explanation for the ``missing electrons'' that would have delocalized into the TMD layers, but also hints at potentially fragile magnetism against low-energy chemical, electrical, and temperature tuning.
\\

\begin{figure}
    \includegraphics[width=0.9 \columnwidth,clip,angle =0]{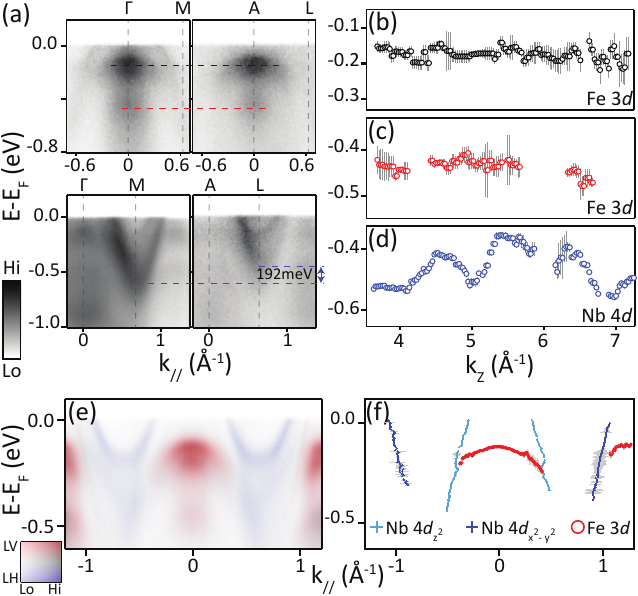}
    \caption{\label{fig3} Mixed dimensionality and hybridization between Fe 3$d$ and Nb 4$d$ states.
    (a) Energy-momentum cut along $\Gamma$-$M$-$K_0$ and $A$-$L$-$H_0$ for $x$ = 0.30 collected at LV (upper panel) and LH (lower panel) polarizations, respectively. The black, red and blue dash lines are visual guides to compare band energy positions.
    (b-d) Scatter points tracking $k_z$ dispersion for Fe-derived bands (black and red circles) and Nb-derived bands (blue circles) extracted from EDC fitting results (Fig.~\ref{SI-SI_kzmap}).
    (e) Data reproduced from Fig.~\ref{fig2}~(a) on $x$~= 0.32, highlighting orbital contributions using a matching color scale to Fig.~\ref{fig2}~(d). (f) Fitted trajectory extracted from (e).
    } 
\end{figure}

\noindent
Given that $d_{z^2}$ orbital is expected to extend along $c$ axis and be the primary orbital to bridge adjacent NbS$_2$ layers, we turn to investigate their potential role in facilitating $c$ axis itinerancy and indirect Fe-Fe magnetic exchange. Figure~\ref{fig3} shows the energy-momentum cut along $\Gamma$-$M$ and $A$-$L$ directions, as well as the $k_z$ dispersion for Fe 3$d$ and Nb 4$d$ bands.
Here, the Fe 3$d_{z^2}$ bands show negligible $k_z$ dispersion (Fig.~\ref{fig3}(b)(c)), confirming the localized nature of low-energy Fe states along all directions.
In contrast, the Nb 4$d$ band gains substantial $k_z$ dispersion on the order of $\sim200$ meV compared to mere $\sim35$ meV in undoped NbS$_2$ (extracted from Fig. 1 (c) of~\cite{NbS2ARPES}).\\

\noindent
The enhanced $c$-axis dispersion of the Nb $4d$ states signifies strong Fe–Nb hybridization, dominated by a direct $\sigma$-overlap between Fe $3d_{z^2}$ and Nb $4d_{z^2}$ orbitals~\cite{NbS2ARPES}, and consistent with their axial alignment along [001].
We visualize the energetic consequence of this interaction in Fig.~\ref{fig3}(e,f) as a distinctive `kink' in the Nb band at $E_B \approx -0.2$ eV, coinciding with its intersection with the Fe $3d_{z^2}$ band.
Crucially, this orbital overlap provides the microscopic pathway for strong, itinerant-mediated RKKY exchange~\cite{RKKY1997,RKKYFeTaS,RKKYKittel}, offering a direct energetic measure of the interaction strength between Fe local moment and NbS$_2$ itinerant electrons~\cite{kondoreview,kondorkky_1,kondorkky_2}.
Such three-dimensional connectivity challenges the strict 2D classification often applied to intercalated TMDs~\cite{parkin1980,friend1977,TMD1968,TMD1971,CrTaS,CoTaS}, distinguishing them from their pristine hosts~\cite{borisenko2009,rahn2012,NbS2ARPES}.
\\

\noindent
Next, to understand the acute transition in charge, magnetic, and transport properties across $x_c$, we turn to the doping dependent electronic structure. Surprisingly, the low-energy Fe 3$d$ states suddenly lose spectral weight above $x_c$ (Fig.~\ref{fig2} (a) and (e), see Fig.~\ref{SI-SI_EDC}~(d) for spectral normalization procedures~\cite{SI}).
Fe site-disorder effect is minimal and smoothly evolves across $x_c$ evidenced by high-resolution synchrotron XRD (Fig.~\ref{SI-SI_XRD}~\cite{SI}). Furthermore, we explicitly exclude micron-scale surface termination inhomogeneity — previously suggested as a potential confounding factor~\cite{EdwaresFNS}—through comprehensive micro-XPS (Fig.~\ref{SI-SI_XPS}~\cite{SI}), real-space micro-ARPES mapping (Fig.~\ref{SI-SI_realspace}~\cite{SI}), as well as the robust persistence of these phenomena over a wide range of photon energies probing deeper into the bulk (Fig.~\ref{fig3}, Fig.~\ref{SI-SI_kzmap}).
Notably, no discernible Fermi surface folding is seen with the magnetic and charge orders, ruling out a possible low-energy Fermi surface nesting scenario (Fig.~\ref{SI-SI_Tdep}~\cite{SI}) \cite{CO-PRL}.
\\

\noindent
Taken together, these findings provide compelling evidence that the abrupt electronic reconstruction may originate from a correlation-driven breakdown of the Fe $3d$ quasiparticles, given quasiparticles are considered well defined only when their energy widths are smaller than their binding energy ($\Delta E/E_B<1$) (see Fig.~\ref{SI-SI_EDC}~(e) for detailed analysis~\cite{SI}).
This sudden decoherence also effectively weakens the $c$-axis Fe $3d_{z^2}$-Nb $4d_{z^2}$ hybridization, suggesting a collapse of the effective $c$-axis hopping channel $t_{\text{Fe-Nb}}$ above $x_c$. This in turn leads to a relative reduction of the out-of-plane Heisenberg exchange parameter $\left|J_2^\prime\right|$ compared to in-plane parameters.
To leading order, such a suppression naturally drives the system from the stripe-ordered ($\left|J_2^\prime\right|>\left|J_2\right|+2\left|J_3^\prime\right|$) to the zigzag-ordered regime ($\left|J_2^\prime\right|<\left|J_2\right|+2\left|J_3^\prime\right|$) [Fig.~\ref{fig1}~(b)]~\cite{sw_prx2022}.
 This scenario is corroborated by spin-resolved DFT calculations, which show that while increasing the effective Hubbard $U$ parameter, a measure of on-site correlation, suppresses all exchange couplings, it preferentially reduces the ratio $\left|J_2^\prime\right|/\left|J_2\right|$ to stabilize the zigzag ground state~\cite{weber2021}.
Interestingly, this mechanism echoes the fragile low-energy hybridization suppressed by the orbital-selective Mott transition (OSMT) of the Fe 3$d_{xy}$ band in iron-based superconductors
\cite{YiFeSe2015,MingYiFeSe,FeSCreview}, where the OSMT strongly reshapes both the local moment and spin density wave tendencies~\cite{FeSCMott_1,FeSCMott_2}. Here, however, the itinerant carriers originate mainly from TMD orbitals, hence the Hund’s coupling effect on Fe is weak and the dominant source of long-range magnetic exchange is via the Fe–Nb $d_{z^2}$ band hybridization.
\\



\begin{figure}
  \includegraphics[width=0.9\columnwidth,clip,angle =0]{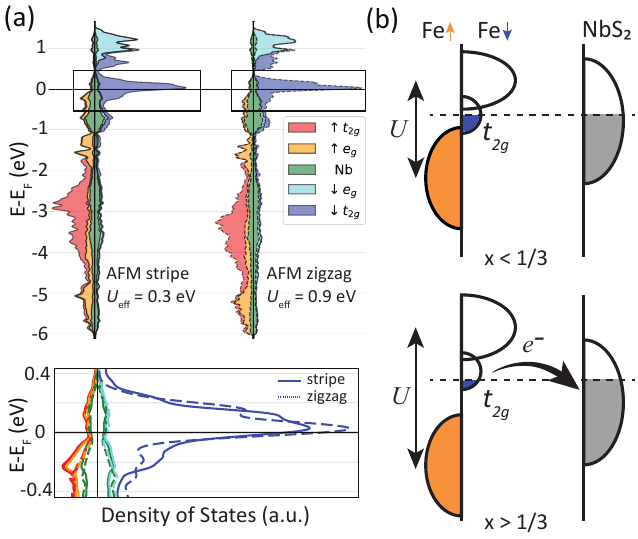}
  \caption{\label{fig4}
  (a) Projected electronic density of states (DOS) of the Fe spin sublattices in Fe$_{1/3}$NbS$_2$ for AFM stripe (left) and AFM zigzag (right) magnetic phases as computed by DFT. The Fe 3$d$ orbitals are further projected by their \ttg \ and \eg \ irreducible representation characters as well as their spin channels of spin-up (bonding) and spin-down (anti-bonding) respectively (indicated by different colors).
  (b) Orbital diagrams depicting the Fe 3$d$ band splitting and occupation below and above the critical doping boundary $x_c = 1/3$.
  }  
\end{figure}

\noindent
To understand the electronic structure associated with the stripe and zigzag phases, we perform spin-polarized DFT calculations for stoichiometric Fe$_{1/3}$NbS$_2$ with the generalized gradient approximation of Perdew, Burke, and Ernzerhof (PBE) functional~\cite{PBE} and a Hubbard $U$ correction~\cite{HubbardU} acting on the Fe $d$ electrons. The Vienna Ab initio simulation package (VASP)~\cite{VASP} is used for all calculations; further computational details are provided in SI~\cite{SI}.
Following prior calculations on Fe$_{1/3}$NbS$_2$~\cite{sw_prx2022,weber2021}, we employ a $U$ of 0.3~eV for the AFM stripe phase ($x<x_c$) and 0.9~eV ($x>x_c$) for the AFM zigzag phase (see Tab.~\ref{SI-tab:DFTenergies} for detailed ground-state energy comparison between different magnetic ordering types~\cite{SI}). These choices lead to computed magnetic moments on the Fe sites of 2.86 and 2.89 $\mu_B$ for stripe and zigzag phases, respectively (see Tab.~\ref{SI-tab:DFTmoments} for detailed comparison of magnetic moments in different ground-states~\cite{SI}), in agreement with prior reports~\cite{FeNbS_CW,sw_prx2022,SAITOH2005}.
\\

\noindent
Figure~\ref{fig4} (a) plots the computed partial density of states (PDOS) for one of the Fe spin sublattices in the stripe (zigzag) antiferromagnetic phase in the left (right) sub-panel with Nb-projected states overlaid.
The pseudo-octahedral coordination about each Fe cation motivates projecting the PDOS onto \ttg \ and \eg \ 3$d$ molecular orbitals, revealing that the strong peak in the DOS near the Fermi level is dominantly of Fe 3$d$ \ttg \ character,
corroborating the interpretation of ARPES data near $x_c$ in Fig.~\ref{fig2}~(a).
The full occupation of the Fe 3$d$ \eg \ bonding orbitals and the partial filling of the Fe 3$d$ anti-bonding \ttg \ orbitals support a high-spin Fe(II) oxidation state consistent with our computed magnetic moments and prior neutron scattering studies~\cite{sw_prx2022}.
Further, the bonding Fe \eg \ 3$d$ states below the Fermi energy divide into two broad levels separated by $\sim2$~eV.
This subsequent splitting of the occupied \eg \ level is evidence of a crystal field associated with the Fe site’s lower $C_{3v}$ global point group symmetry~\cite{XieCrystalField} and reflects significant hybridization of the Fe and Nb states.
Strong Fe-Nb hybridization
is evident from the greater broadening of the occupied compared to the unoccupied Fe 3$d$ states, and from their energy overlap with the occupied Nb 4$d$ states in the Nb-projected PDOS shown in Fig.~\ref{fig4}~(a).\\


\noindent
Notably, the partially-occupied \ttg \ states residing at the Fermi energy ($E_F$) are computed to differ for the two phases, with substantial reorganization of these antibonding 3$d$ states in the zigzag phase relative to stripe.
As shown in the zoomed-in view between -400~meV and 400~meV of the PDOS plot in Fig.~\ref{fig4}~(a), in the transition from the stripe to the zigzag phase, this \ttg \ level is pushed above $E_F$, while the PDOS tail falls 400~meV below $E_F$, qualitatively consistent with ARPES observation for $x > x_c$ (Fig.~\ref{fig2}, see Fig.~\ref{SI-SI_bands} for the calculated unfolded band structure).
More drastic differences in the PDOS of the occupied \ttg \ orbitals further below $E_F$ between the two phases are also observed in our calculations.
The schematic in Fig.~\ref{fig4}~(b) summarizes the electronic structure evolution across $x_c$.
As $x$ increases above 1/3, the zigzag phase is favored, where the increased $U$ depopulates these \ttg \ states and causes a decoherence of occupied Fe states above $x_c$, leading to more electrons returning to the Nb 4$d$ states.
This occupation change of Fe \ttg \ states is consistent with the observation in Fig.~
\ref{fig2}~(b), as well as a slightly enhanced Fe pre-$L_3$ edge peak for $x$ = 0.36 in XAS measurements (Fig.~\ref{SI-SI_XAS}~\cite{SI}) and the slightly increased ordered moment size due to the removal of anti-bonding Fe electrons for $x > x_c$ seen in neutron scattering~\cite{sw_prx2022}.
\\

\noindent
To conclude, by isolating the moment-carrying Fe states near $E_F$, we have revealed that the sudden collapse of local-itinerant hybridization across $x_c$ dictates the effective exchange $J$ and drives the magnetic phase transition.
This correlation-driven electronic reconstruction distinguishes Fe-intercalated TMDs from isostructural systems exhibiting continuous evolution, offering a clear explanation for the sharp phase boundary at $x_c = 1/3$.
Methodologically, we demonstrate that ARPES can effectively reveal $J_{\text{eff}}$ in metallic magnets by directly visualizing orbital-specific hybridization via identification of interacting orbitals and the hybridization strengths, offering a powerful complementary route to spin-wave modeling of inelastic neutron or x-ray scattering spectra in metallic magnets.
Looking forward, atomically resolved probes such as scanning tunneling microscope/spectroscopy (STM/S), and surface-specific magnetic probes such as XPEEM, may help shed light on the atomic origin of the sudden change in the correlation strength in Fe across $x_c$, potentially by mapping local Fe vacancy or interstitial disorders to specific magnetization states.

\section{Acknowledgments}
\noindent
We thank Eduardo H. da Silva Neto, Pranab Kumar-Nag, Xinze Yang, Sophie Weber, Tianyu Zhu, Ke Liao, Ming Yi, and Ruihua He for helpful discussions.
Work at Yale University was funded by the U.S. Air Force Office of Scientific Research under Award No. FA9550-24-1-0048.
W. L. acknowledges support from James Kouvel Fellowship.
Work at the University of California, Berkeley and Lawrence Berkeley National Laboratory was funded by the U.S. DOE, Office of Science, Office of Basic Energy Sciences, Materials Sciences and Engineering Division under Contract No. DE-AC02-05CH11231 (Quantum Materials Program KC2202 and Theory of Materials FWP). Theoretical calculations were carried out using the National Energy Research Scientific Computing Center (NERSC).
Work at the University of California, San Diego was supported by the National Science Foundation under Grant No. DMR-2145080.
Bulk crystal growth of NbS$_2$ materials was supported by the 2DCC-MIP under NSF cooperative agreement DMR-2039351.
Use of the Stanford Synchrotron Radiation Lightsource, SLAC National Accelerator Laboratory, is supported by the U.S. Department of Energy, Office of Science, Office of Basic Energy Sciences under Contract No. DE-AC02-76SF00515.
This research used resources of the National Synchrotron Light Source II, operated under Contract No. DE-SC0012704. The authors also acknowledge the Brookhaven National Laboratory-Yale partner user agreement PU-313536.
We acknowledge Yale West Campus Materials Characterization Core.
This work is based on research conducted at the Center for High-Energy X-ray Sciences (CHEXS), which is supported by the National Science Foundation (BIO, ENG and MPS Directorates) under award DMR-2342336.
This research used resources of the Advanced Light Source, a US DOE Office of Science User Facility under Contract No. DE-AC02-05CH11231.
This research used resources of the Advanced Photon Source, a U.S. Department of Energy (DOE) Office of Science User Facility operated for the DOE Office of Science by Argonne National Laboratory under Contract No. DE-AC02-06CH11357.

\bibliography{bibfile.bib}

\AddToHook{enddocument/afteraux}{%
\immediate\write18{
cp output.aux main-r2-clean.aux
}%
}
\end{document}